\documentclass[twocolumn,english,aps,pra,floatfix]{revtex4}
\usepackage[T1]{fontenc}
\usepackage[latin9]{inputenc}
\usepackage{color}
\usepackage{amsmath}
\usepackage{amssymb}
\usepackage{graphicx}
\usepackage{natbib}
\usepackage{braket}
\usepackage{psfrag}
\usepackage{tikz}
\usepackage{array}
\usepackage[normalem]{ulem}

\usepackage[colorlinks=true,urlcolor=blue,citecolor=blue,linkcolor=blue]{hyperref}

\usepackage{babel}

\newcolumntype{x}[1]{>{\centering\arraybackslash\hspace{12pt}}p{#1}}

\begin{document}

\title{Entanglement generation in superconducting qubits using holonomic operations}

\author{D. J. Egger}
\affiliation{IBM Research GmbH, Zurich Research Laboratory, S\"aumerstrasse 4, 8803 R\"uschlikon, Switzerland}
\author{M. Ganzhorn}
\affiliation{IBM Research GmbH, Zurich Research Laboratory, S\"aumerstrasse 4, 8803 R\"uschlikon, Switzerland}
\author{G. Salis}
\affiliation{IBM Research GmbH, Zurich Research Laboratory, S\"aumerstrasse 4, 8803 R\"uschlikon, Switzerland}
\author{A. Fuhrer}
\affiliation{IBM Research GmbH, Zurich Research Laboratory, S\"aumerstrasse 4, 8803 R\"uschlikon, Switzerland}
\author{P. M\"uller}
\affiliation{IBM Research GmbH, Zurich Research Laboratory, S\"aumerstrasse 4, 8803 R\"uschlikon, Switzerland}
\author{P. Kl. Barkoutsos}
\affiliation{IBM Research GmbH, Zurich Research Laboratory, S\"aumerstrasse 4, 8803 R\"uschlikon, Switzerland}
\author{N. Moll}
\affiliation{IBM Research GmbH, Zurich Research Laboratory, S\"aumerstrasse 4, 8803 R\"uschlikon, Switzerland}
\author{I. Tavernelli}
\affiliation{IBM Research GmbH, Zurich Research Laboratory, S\"aumerstrasse 4, 8803 R\"uschlikon, Switzerland}
\author{S. Filipp}
\affiliation{IBM Research GmbH, Zurich Research Laboratory, S\"aumerstrasse 4, 8803 R\"uschlikon, Switzerland}

\date{\today}

\begin{abstract}
We investigate a non-adiabatic holonomic operation that enables us to entangle two fixed-frequency superconducting transmon qubits attached to a common bus resonator. Two coherent microwave tones are applied simultaneously to the two qubits and drive transitions between the first excited resonator state and the second excited state of each qubit. The cyclic evolution within this effective 3-level $\Lambda$-system gives rise to a holonomic operation entangling the two qubits. Two-qubit states with 95\% fidelity, limited mainly by charge-noise of the current device, are created within $213~\rm{ns}$. This scheme is a step toward implementing a SWAP-type gate directly in an all-microwave controlled hardware platform. By extending the available set of two-qubit operations in the fixed-frequency qubit architecture, the proposed scheme may find applications in near-term quantum applications using variational algorithms to efficiently create problem-specific trial states.
\end{abstract}

\date{\today}

\maketitle

\section{Introduction}

Superconducting qubits are one of the leading candidates to build a quantum computer \cite{Devoret2013}. They are fabricated with conventional micro- and nanofabrication techniques \cite{Oliver2013b, Chang2013} and are controlled using standard microwave instrumentation. The fixed-frequency transmon is a specific version of a superconducting qubit that is not sensitive to flux \cite{Koch2007}. This improves the qubit coherence time at the cost of controllability. Fixed-frequency qubits can be coupled dispersivly through coupling elements such as coplanar wave-guide resonators \cite{Blais2007, Majer2007}. Two-qubit operations are then activated by applying coherent microwave signals to generate, e.g., a controlled-NOT operation using the cross-resonance gate \cite{Rigetti2010, Sheldon2016b}.

In this paper we demonstrate an alternative all-microwave entangling scheme in which we simultaneously drive transitions between a resonator and two fixed-frequency qubits \cite{Zeytinoglu2015}. This scheme is based on a holonomy emerging in a driven $\Lambda$-type system with three energy levels \cite{Sjoqvist2012, Abdumalikov2013, Gasparinetti2016}. Such quantum operations have attracted attention because they may be exploited for holonomic quantum computing based on non-abelian geometric phases created by steering the system along a closed loop in Hilbert space \cite{Zanardi1999, Sjoqvist2015}. Holonomic quantum computing may benefit from the robustness of geometric phases to certain types of errors \cite{Blais2003, Whitney2003, De2003, Carollo2003a, Solinas2004, Leek2007, Filipp2009a, Cucchietti2010, Berger2013, Wu2013a, Berger2015, Zheng2016, Johansson2012}. While in theory holonomic adiabatic gates can reach high fidelities in the limit of long gate duration \cite{Kamleitner2011, Johansson2012}, decoherence in real devices severly limits the achievable gate fidelities. A particular challenge is to realize holonomic two-qubit operations. A few experimental demonstrations exist but operate either on different degrees of freedom \cite{Zu2014} or use a trotterized approach \cite{Feng2013a}. Here, we present a direct realization of a non-adiabatic two-qubit holonomic operation that entangles two superconducting qubits in a scalable architecture \cite{QuantumExperience}. Moreover, in contrast to the cross-resonance gate which realizes a CNOT primitive, the holonomy results in an exchange-type operation which may be useful to reduce the circuit depth in various quantum algorithms \cite{Barkoutsos2018}.

This paper is structured as follows. Section \ref{Sec:setup} introduces the holonomic operation and the experimental setup. Section \ref{Sec:cal} discusses the various steps needed to calibrate the holonomic operation. Experimental results and simulations are discussed in sections \ref{Sec:experiments} and \ref{Sec:sim}, respectively.

\section{Setup and holonomic operation \label{Sec:setup}}

\begin{center}
\begin{figure}[!ht]
\includegraphics[width=0.48\textwidth, clip, trim=20 45 500 0]{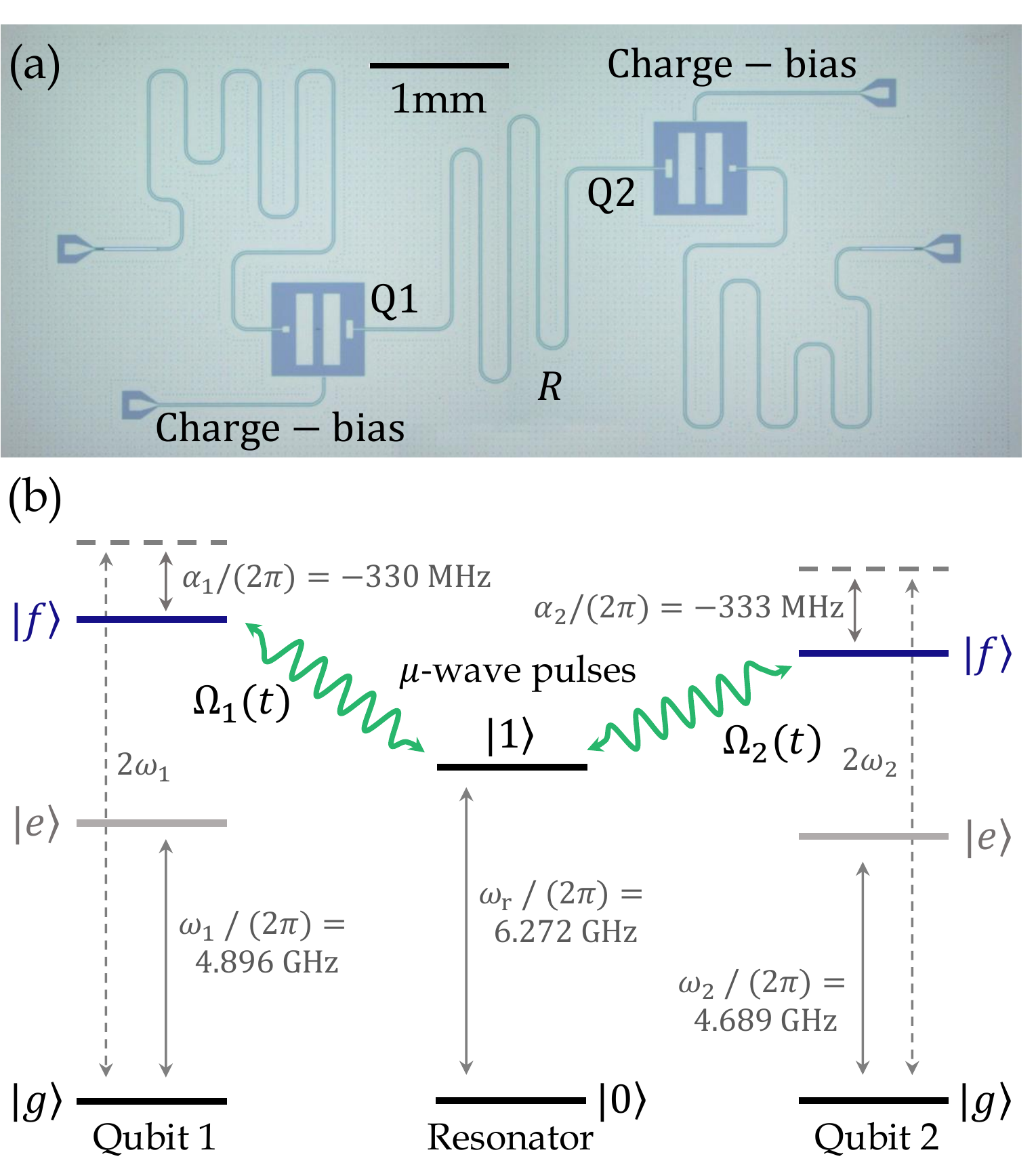}
\caption{\label{Fig:levels} (a) Micrograph of the superconducting qubit chip with two transmons (Q1 and Q2) connected via a coplanar wave-guide resonator (R). The transmon-style qubits are individually addressed through capacitively coupled charge-bias lines and measured using coplanar wave-guide readout resonators. (b) Level diagram of the two qubits with energy levels $\ket{g}$, $\ket{e}$ and $\ket{f}$ connected to a resonator  with lowest lying states $\ket{0}$ and $\ket{1}$. The holonomic operation is created by microwave drives $\Omega_{1,2}(t)$ between the $\ket{f}$ state of each qubit and the $\ket{1}$ state of the resonator.}
\end{figure}
\end{center}

The system is made up of two fixed-frequency superconducting qubits coupled to a common coplanar wave-guide resonator, see Fig.\ \ref{Fig:levels}(a). The coupling strengths are $g_1/(2\pi)=156~\rm{MHz}$ and $g_2/(2\pi)=196~\rm{MHz}$ and the resonator frequency is $\omega_\text{r}/(2\pi)=6.272~\rm{GHz}$. The Hamiltonian describing the system is
\begin{align} \notag
\hat{H} = \omega_\text{r}\hat a^\dagger\hat a+\sum\limits_{i=1,2}\omega_i\hat b_i^\dagger\hat b_i^{\phantom{\dagger}}+\frac{\alpha_i}{2}\hat b_i^\dagger\hat b_i^\dagger\hat b_i^{\phantom{\dagger}}\hat b_i^{\phantom{\dagger}}  \\
+ g_i\left(\hat b_i^\dagger\hat a+\hat b_i^{\phantom{\dagger}}\hat a^\dagger\right)+\Omega_i(t)\left(\hat b_i^\dagger+\hat b_i^{\phantom{\dagger}}\right). \label{Eqn:H}
\end{align}
Here $\hbar=1$, $\hat b_i$ ($\hat b_i^\dagger$) is the lowering (raising) operator of qubit $i$ whilst $\hat a$ ($\hat a^\dagger$) is the resonator lowering (raising) operator. The qubit frequencies and anharmonicities are $\omega_1/(2\pi)=4.896~\rm{GHz}$ and $\omega_2/(2\pi)=4.689~\rm{GHz}$ and $\alpha_1/(2\pi)=-330~\rm{MHz}$ and $\alpha_2/(2\pi)=-333~\rm{MHz}$, respectively. The system is operated in the dispersive regime $|\Delta_i|=|\omega_i-\omega_\text{r}|\gg g_i$ to avoid direct qubit-resonator excitation transfer. The $T_1$ times of qubit one (Q1), qubit two (Q2) and the coupling resonator are $42~\mu\rm{s}$, $56~\mu\rm{s}$ and $7~\mu\rm{s}$, respectively. Each qubit is driven by a microwave signal with a time-dependent amplitude $\Omega_i(t)$ applied via individual capacitively coupled charge bias lines.

\subsection*{Entanglement with holonomic operations \label{Sec:Theory}}
In our experiments we use the first three qubit states $\ket{g}$, $\ket{e}$ and $\ket{f}$ here listed in order of increasing energy. The two-qubit operation described in this paper is realized by simultaneously driving the $\ket{f0}$ to $\ket{g1}$ transition of each qubit where $\ket{0}$ is the ground state and $\ket{1}$ is the excited state of the resonator, see Fig.\ \ref{Fig:levels}(b). These qubit-resonator transitions, as well as the single qubit transitions, are controlled by applying the microwave drive $\Omega_i(t) = \lambda_i\eta_i(t) \cos(\omega_{\text{d},i} t + \varphi_i)$ with adjustable frequency $\omega_{\text{d},i}$, phase $\varphi_i$, and pulse envelope $\eta_i(t)$ to each qubit. The complex scalars $\lambda_i$ are the relevant scaling factors for the two-qubit holonomic operation.

Applying a drive on qubit $i$ at frequency $\omega_{\text{d},i}=\omega_i$ ($\omega_{\text{d},i}=\omega_i+\alpha_i$) creates a rotation $X_{\beta_i}^{g\to e}$ ($X_{\beta_i}^{e\to f}$) around the x-axis of the $\{\ket{g},\ket{e}\}$ ($\{\ket{e},\ket{f}\}$) Bloch sphere with angle $\int\eta_i(t)\text{d}t=\beta_i$  \cite{Motzoi2009, Bianchetti2010}. We set $\lambda_i=1$ for single qubit operations. A different rotation axis in the equatorial plane can be selected by changing the phase $\varphi_i$ of the drive. Similarly, applying a drive on qubit $i$ at the difference frequency between the $\ket{f}$ state of qubit $i$ and the excited resonator state $\ket{1}$, i.e.\ $\omega_{\text{d},i}=2\omega_i+\alpha_i-\omega_\text{r}$, activates induced Jaynes-Cummings-type vacuum-Rabi oscillations  between these states entangling the qubit and the resonator \cite{Zeytinoglu2015}. This can be seen from a perturbation theory argument considering only qubit $i$, the resonator and the drive, which results in the effective Hamiltonian \cite{Pechal2014, Zeytinoglu2015}
\begin{align} \notag
\hat H_{\text{eff},i}=&\Delta^{(i)}_\text{f0}\ket{f0}\!\bra{f0}
+\tilde{g}_i\ket{f0}\!\bra{g1}+\text{h.c.}
\end{align}
The qubit state $\ket{e}$ is far detuned and can be ignored. The ac-Stark shift $\Delta^{(i)}_\text{f0}$ is to leading order quadratic in the drive strength. The drive frequency can be set to compensate for this ac-Stark shift so that the states $\ket{f0}$ and $\ket{g1}$ form a degenerate subspace allowing for a coherent population transfer between qubit and resonator. The effective coupling strength between $\ket{f0}$ and $\ket{g1}$ is \cite{Zeytinoglu2015}
\begin{align} \label{Eqn:geff}
\tilde{g}_i(t)=\frac{g_i\alpha_i\lambda_i\eta_i(t)}{\sqrt{2}\Delta_i(\Delta_i+\alpha_i)}.
\end{align}
The rate of the $\ket{f0}\leftrightarrow\ket{g1}$ microwave activated transition decreases with qubit-resonator detuning $\Delta_i$ but can be compensated by stronger driving. Rabi rates of $10~\rm{MHz}$ have been reported \cite{Zeytinoglu2015}.

Simultaneously applying both drives, illustrated in Fig.\ \ref{Fig:levels}(b), creates a degenerate subspace spanned by $\{\ket{fg0},\ket{gf0},\ket{gg1}\}$. Again, perturbation theory gives the effective Hamiltonian
\begin{align} \label{Eq:Heff}
\hat H_\text{eff}(t)=\,&\Delta_\text{fg0}\ket{fg0}\!\bra{fg0}
+\Delta_\text{gf0}\ket{gf0}\!\bra{gf0} \\
&+\tilde{g}_1(t)\ket{fg0}\!\bra{gg1}+\tilde{g}_2(t)\ket{gf0}\!\bra{gg1}+\text{h.c.} \notag
\end{align}
with the effective coupling strengths given by Eq.\ (\ref{Eqn:geff}). The ac-Stark shifts $\Delta_\text{fg0}$ and $\Delta_\text{gf0}$ under this two-tone drive, henceforth named cross ac-Stark shifts, may differ in experiment from the ac-Stark shifts $\Delta_\text{f0}^{(i)}$ under the single tone drive, see Sec.\ \ref{Sec:cal}. The cross ac-Stark shifts are removed by approprietly selecting the drive frequencies. To create a holonomic operation and to avoid transitions into dark states \cite{Sjoqvist2012, Fleischhauer1996}, the Rabi rates at each point in time must be equal up to the constant scaling parameters $\lambda_i$, i.e. 
\begin{align} \label{Eqn:eff_rates_equal}
\frac{\tilde{g}_1(t)}{\lambda_1}=\frac{\tilde{g}_2(t)}{\lambda_2}=\tilde g(t)~\forall t.
\end{align}
Doing so allows us to write the Hamiltonian in Eq.\ (\ref{Eq:Heff}) as the product of a time dependent constant and a time independent operator
\begin{align} \label{Heff_no_stark}
\hat H_\text{eff}'(t)=\tilde g(t)\left(\lambda_1\ket{fg0}\!\bra{gg1}+\lambda_2\ket{gf0}\!\bra{gg1}+\text{h.c.}\right).
\end{align}
Note that Eq.\ (\ref{Eqn:eff_rates_equal}) and (\ref{Eqn:geff}) imply that both pulses have the same envelope $\eta_1(t)=\eta_2(t)$. Under $\hat H_\text{eff}'$ the system evolves according to 
\begin{align} \notag
\ket{\psi_j(t)}=\exp\left(-i\int_0^t\hat H_\text{eff}'(\tau)\mathrm{d}\tau\right)\ket{\psi_j(0)},
\end{align}
where $\ket{\psi_{j}(0)}\in\{\ket{fg0},\ket{gf0},\ket{gg1}\}$. States that start in the subspace spanned by $\ket{fg0}$ and $\ket{gf0}$ satisfy the parallel transport condition $\braket{\psi_j(t)|\hat H_\text{eff}'(t)|\psi_k(t)}=0$, see Appendix \ref{Sec:parallel}. The evolution is thus purely geometric \cite{Sjoqvist2012, Fleischhauer1996}. If the pulse duration $T$ is chosen such that the evolution is cyclic, i.\ e.\ $\int_0^T\tilde g_i(t)\mathrm{d}t=\pi$, the final state $\ket{\psi_j(T)}$ returns to its initial subspace $\{\ket{fg0},\ket{gf0}\}$, see Appendix \ref{Sec:dynamics}. In this basis the resulting operation can be written as the transformation
\begin{align} \label{Eqn:U}
\hat U =\begin{pmatrix}
 \cos\theta & e^{i\phi}\sin\theta \\
 e^{-i\phi}\sin\theta & -\cos\theta
\end{pmatrix},
\end{align}
where
\begin{equation} \label{Eqn:phiTheta}
\left\{
\begin{aligned}
-e^{i\phi}\tan\frac{\theta}{2}=&\, \frac{\lambda_1}{\lambda_2}, \\
|\lambda_1|^2+|\lambda_2|^2=&\, 1.
\end{aligned}\right.
\end{equation}
The form of $\hat U$ implies that arbitrary rotations between $\ket{fg}$ and $\ket{gf}$ can be created by changing $\lambda_1$ and $\lambda_2$. The rotation angle $\theta$ controls the amount of population transferred between the qubits. The phase difference  between the drives $\phi=\varphi_1-\varphi_2$ changes the relative phase between $\ket{fg}$ and $\ket{gf}$. The magnitudes of the rotation $\theta$ and phase $\phi$ only depend on $\lambda_i$ and are controlled at fixed operation time $T$.

As an example, let $\lambda_1=\lambda_2=1/\sqrt{2}$ and Q1 and Q2 be in the $\ket{f}$ and $\ket{g}$ states, respectively. The drive $\Omega_1(t)$ transfers the population from Q1 to the resonator. Simultaneously, the second drive $\Omega_2(t)$ transfers the population from the resonator to Q2. During this time-evolution the resonator is only partially populated and is completely depleted at the end, as discussed in Sec.\ \ref{Sec:sim}.

\section{Calibration of the Simultaneous two-tone drive \label{Sec:cal}}

\begin{center}
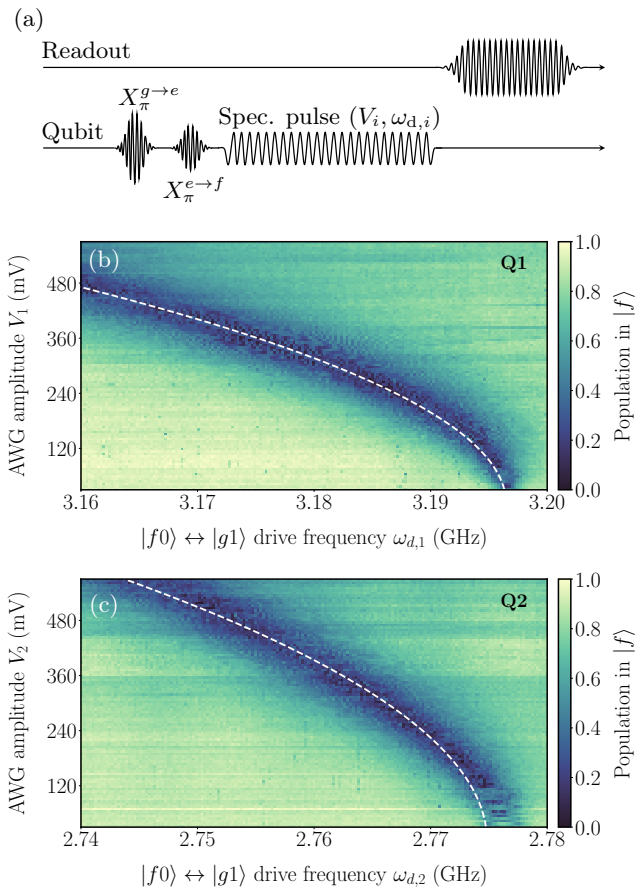
\begin{figure}[!ht]
\begin{tikzpicture}
\node at (0,6.0cm) {\includegraphics[width=0.44\textwidth,clip, trim=0 300 0 0]{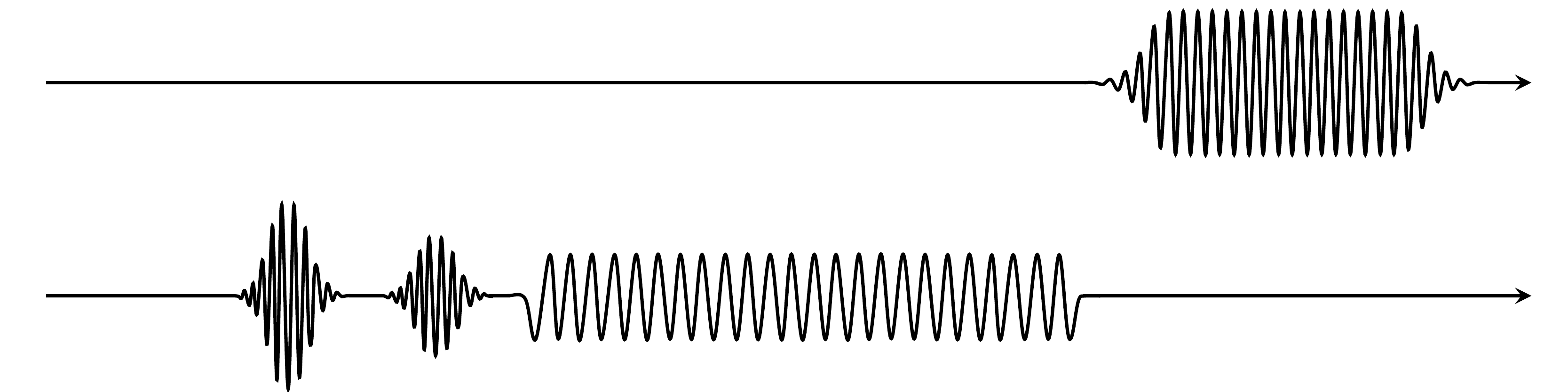}};
\node at (-3.9cm,7.2cm) {(a)};
\node at (-2.3cm,6.15cm) {$X_\pi^{g\to e}$};
\node at (-1.7cm,4.95cm) {${X_\pi^{e\to f}}$};
\node at (0.1cm,5.9cm) {\small Spec. pulse $(V_i, \omega_{\text{d},i})$};
\node[anchor=west] at (-3.85cm,6.8cm) {\small Readout};
\node[anchor=west] at (-3.85cm,5.7cm) {\small Qubit};
\node at (0,0) {\includegraphics[width=0.47\textwidth]{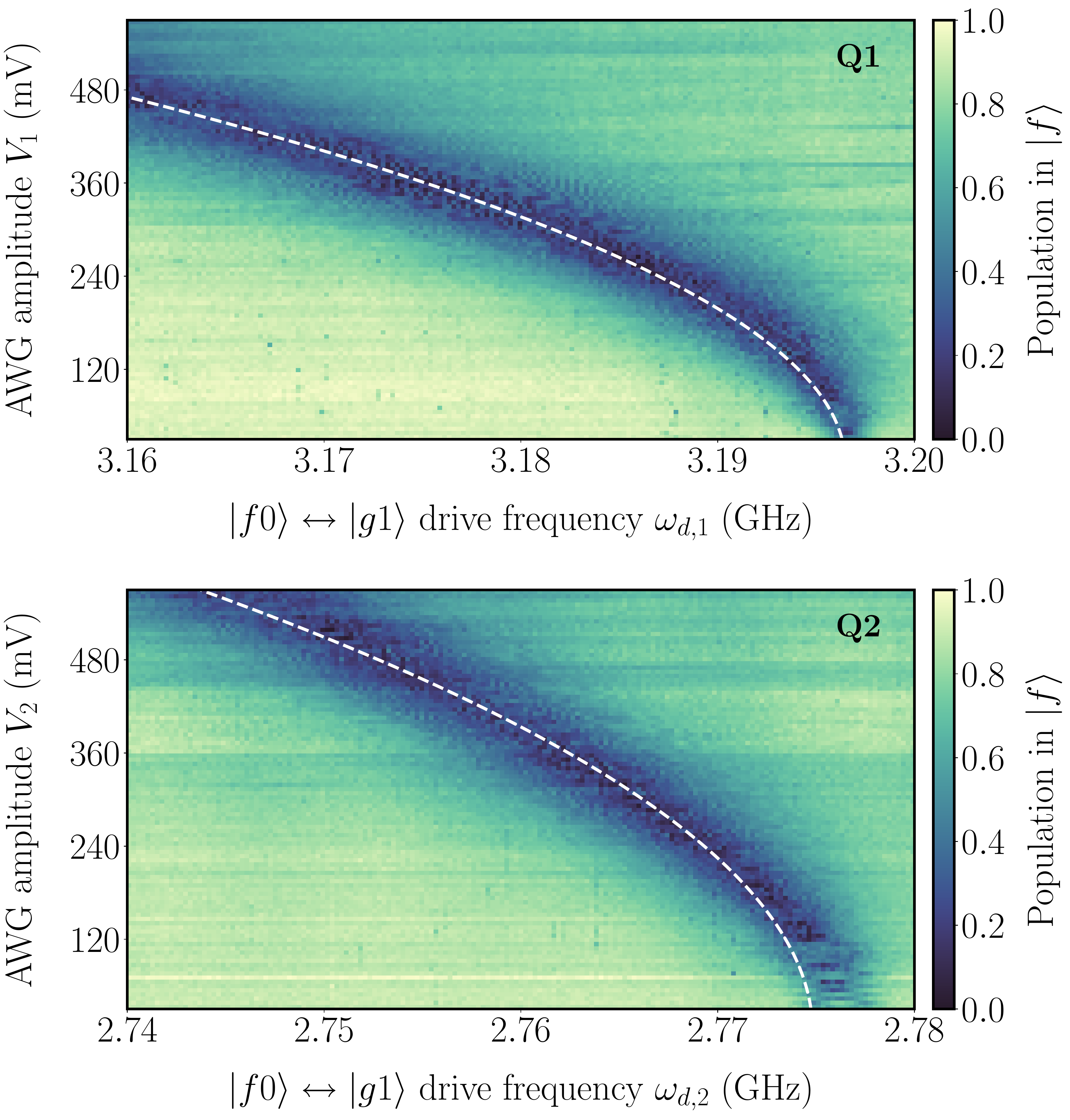}};
\node[white] at (-2.9cm,4.0cm) {(b)};
\node[white] at (-2.9cm,-0.6cm) {(c)};
\end{tikzpicture}
\caption{\label{Fig:StarkShift} (a) Pulse sequence for qubit and readout tone used to calibrate the ac-Stark shifts. (b) and (c) Calibration of the ac-Stark shifts for Q1 and Q2 respectively. The qubit $\ket{f}$ population is measured after a spectroscopic square pulse around the targeted $\ket{f0}\leftrightarrow\ket{g1}$ transition. The y-axis is given as voltage amplitude $V_i$ at the output of the arbitrary waveform generator that drives the respective qubit $i$. For each amplitude the spectrum is fitted to a Lorentzian, see text. The ac-Start shift corrected $\ket{f0}\leftrightarrow\ket{g1}$ frequencies are $\omega_{\text{f0}\leftrightarrow\text{g1}}^{(i)}/(2\pi)=a_iV_1^2+b_i~\rm{GHz}$ with coefficients $\{a_1,b_1\}=\{-0.164~\rm{GHz/V^2},3.196~\rm{GHz}\}$ and $\{a_2,b_2\}=\{-0.094~\rm{GHz/V^2},2.775~\rm{GHz}\}$ for Q1 and Q2, respectively.
}
\end{figure}
\end{center}

\begin{center}
\begin{figure}[!ht]
\includegraphics[width=0.48\textwidth]{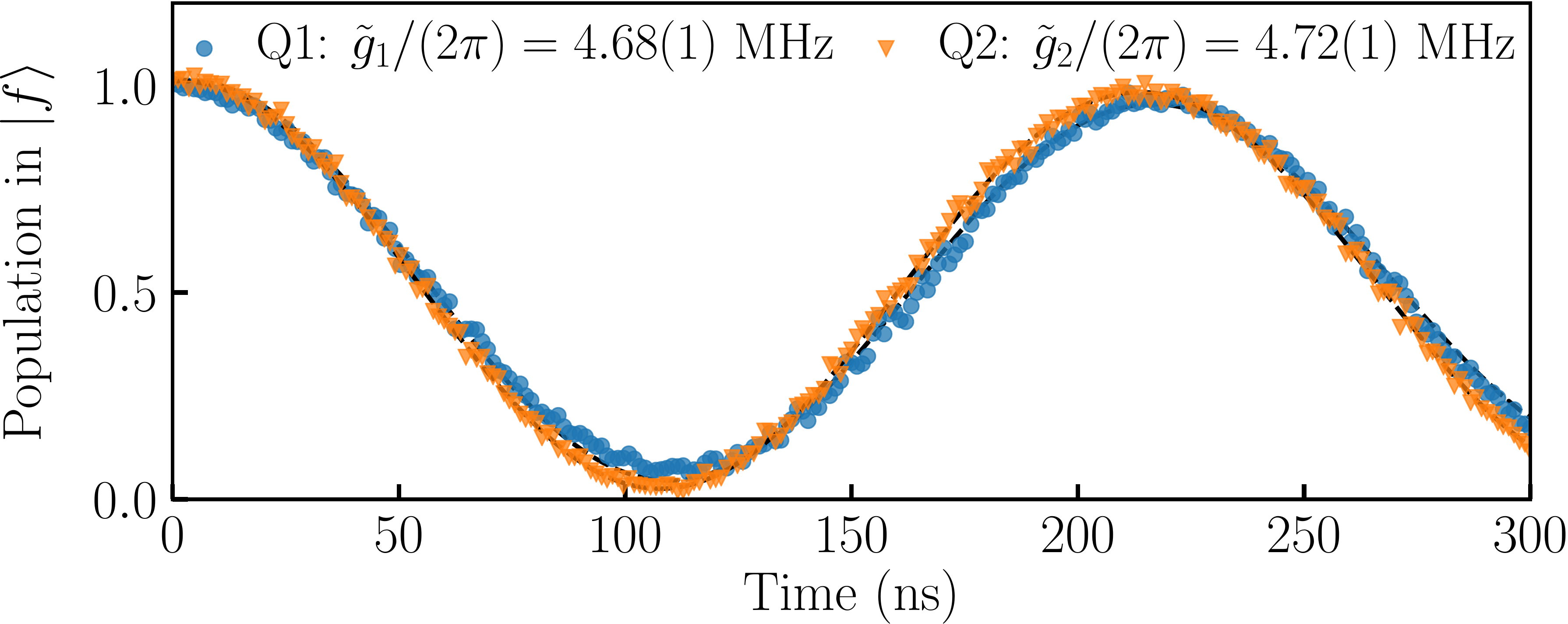}
\caption{\label{Fig:rates} Rabi oscillations between each qubit $\ket{f}$ state and the resonator $\ket{1}$ state. Each qubit is first prepared in $\ket{f}$, then the $\ket{f0}\leftrightarrow\ket{g1}$ drive is applied for a variable time before measuring the qubit. The resulting data set is fitted to a decaying sinusoid to extract the effective coupling strength $\tilde{g}_i$.}
\end{figure}
\end{center}

To realize the holonomic operation $\hat U$ in Eq.\ (\ref{Eqn:U}), the strength, frequency, phase, and duration of each drive must be controlled. In particular, cross ac-Stark shifts occurring under simultaneous driving of both qubit-resonator transitions must be compensated to avoid frequency offsets that reduce the fidelity of the population transfer. To keep all ac-Stark shifts constant during the holonomic operation, square pulse envelopes are used when driving the $\ket{f0}\leftrightarrow\ket{g1}$ transitions. In short, the calibration is accomplished through the following steps.
\begin{enumerate}
 \item[(i)] Measurements of individual ac-Stark shifts when applying a single drive $\Omega_i(t)$ result in a calibration curve of frequency shifts as a function of drive amplitude, see Fig.\ \ref{Fig:StarkShift}.
\item[(ii)] Jaynes-Cummings-type oscillations between $\ket{f0}$ and $\ket{g1}$ are measured to calibrate the individual drive amplitudes with $\lambda_i=1$ such that the induced Rabi rates are equal, i.e.\ $\tilde{g}_1=\tilde{g}_2$, see Fig.\ \ref{Fig:rates}.
\item[(iii)] To calibrate cross ac-Stark shifts, the population transferred to an initially empty target qubit is measured as a function of drive frequency offset  when both drives are applied simultaneously, see Fig.\ \ref{Fig:freq_sweep}(a).
\item[(iv)] Changing the rotation angle $\theta$ requires modifying the drive strengths thus producing different (cross) ac-Stark shifts. Step (iii) is, therefore, repeated for different $\theta$ angles to obtain a $\theta$-dependent calibration curve of cross ac-Stark shifts for each drive, see Fig.\ \ref{Fig:freq_sweep}(b).
\end{enumerate}

In more detail, in step (i) the ac-Stark shift induced by the $\ket{f0}\leftrightarrow\ket{g1}$ drive of each qubit is determined for a range of drive amplitudes. Spectroscopy on the $\ket{f0}\leftrightarrow\ket{g1}$ transition of each qubit yields the $\Delta^{(i)}_\text{f0}(\Omega_i)$ ac-Stark shift calibration curve. In each data set the relevant qubit is prepared in the $\ket{f}$ state by applying a $X_{\pi}^{g\to e}$ $\pi$-pulse followed by a $X_{\pi}^{e\to f}$ $\pi$-pulses. Next, a $10~\mu\rm{s}$ spectroscopic pulse is applied at a fixed amplitude and frequency and the qubit is measured, see Fig.\ \ref{Fig:StarkShift}(a). This pulse sequence is repeated for different amplitudes and frequencies. The resonance line is fitted to a second-order polynomial in drive amplitude shown by the white dashed line in Fig.\ \ref{Fig:StarkShift}(b)-(c). In the limit of zero drive amplitude the frequencies of the $\ket{f0}\leftrightarrow\ket{g1}$ drives are $3.196~\rm{GHz}$ and $2.775~\rm{GHz}$ for Q1 and Q2, respectively. Henceforth, all frequency shifts in the following figures will be referenced to these values.

\begin{center}
\begin{figure}[!ht]
\begin{tikzpicture}
\node at (0,-0.5cm) {\includegraphics[width=0.47\textwidth]{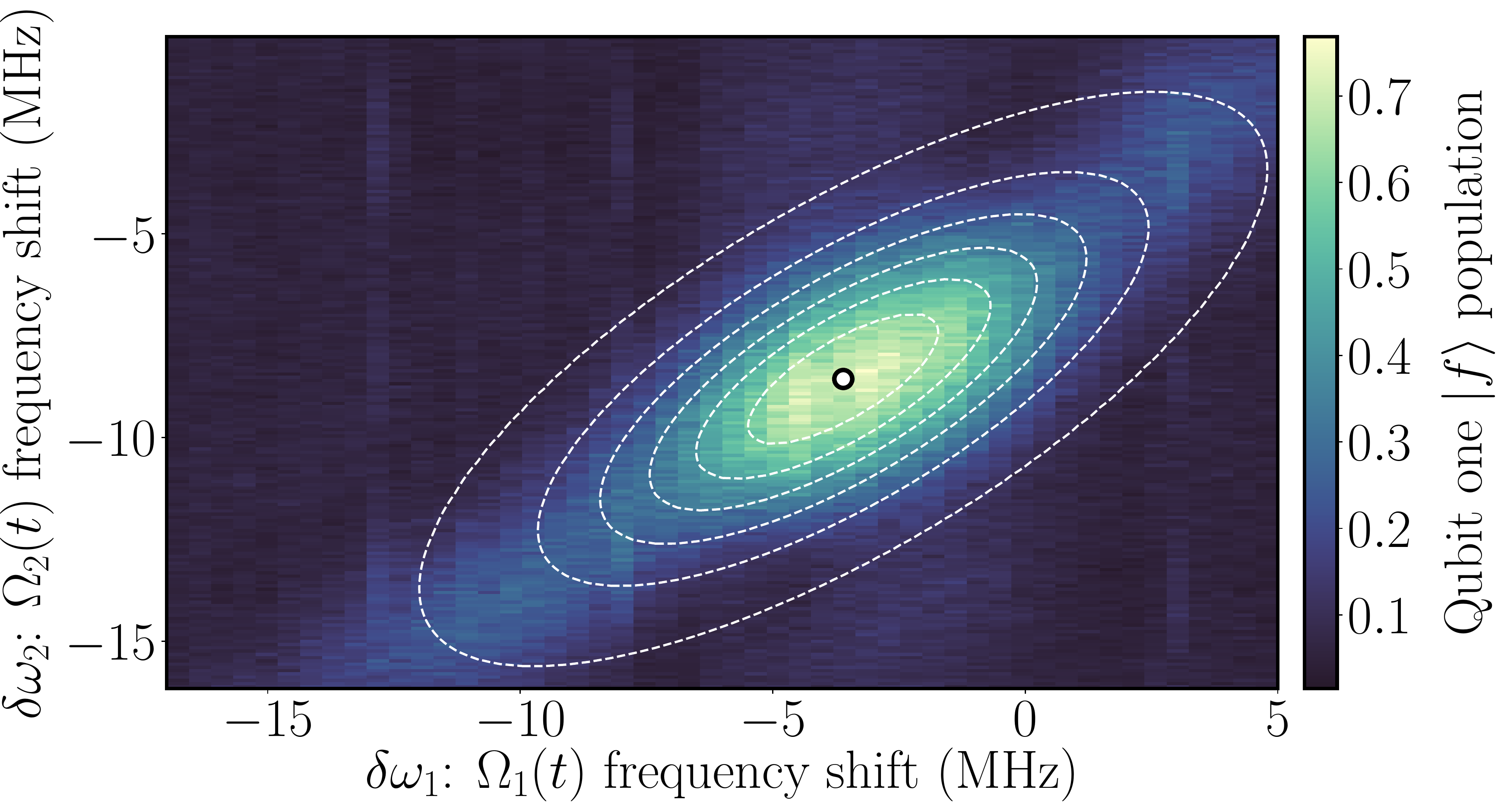}};
\node[white] at (-2.9cm,1.2cm) {(a)};
\node at (0,-5.1cm) {\includegraphics[width=0.47\textwidth]{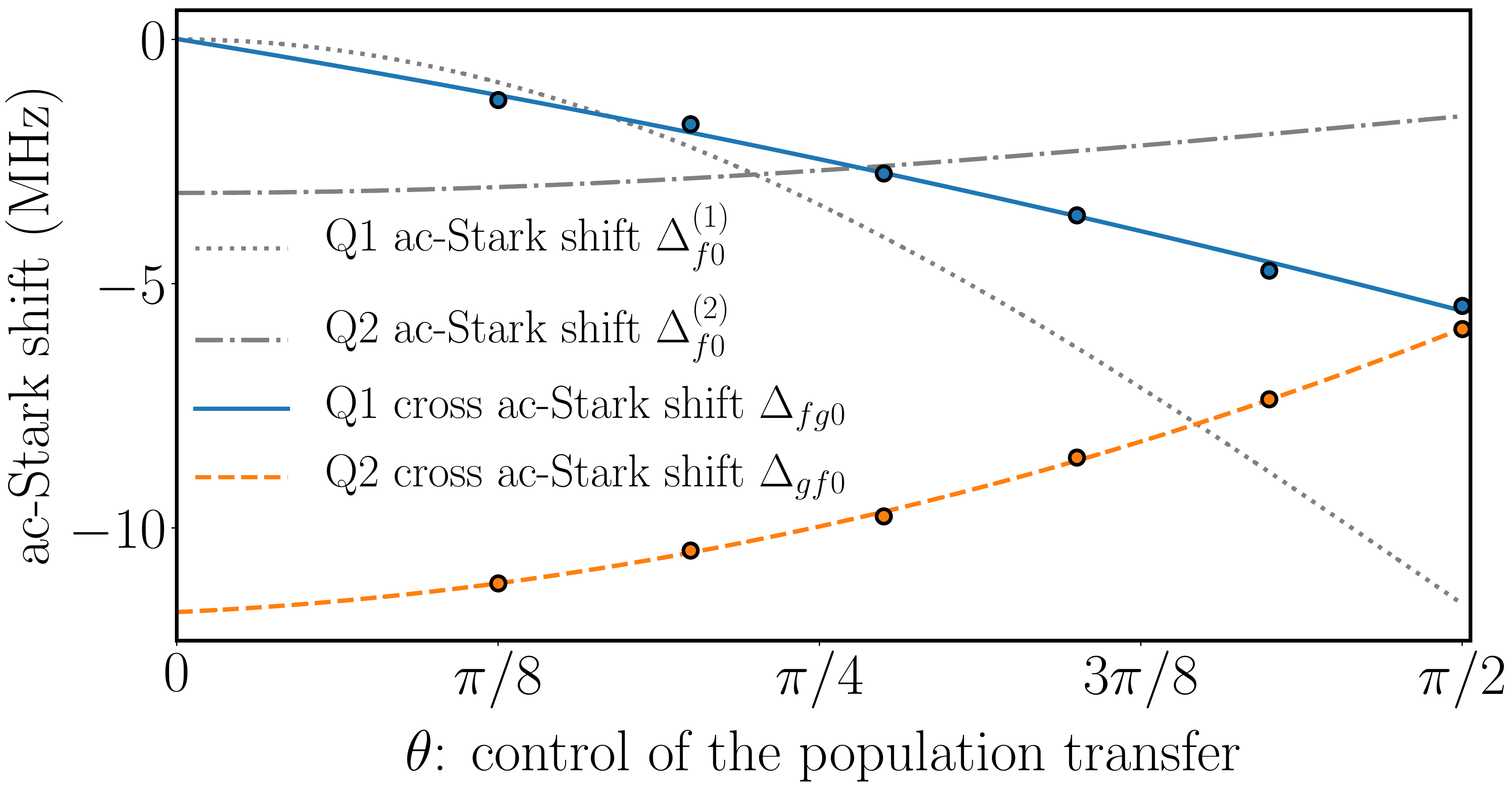}};
\node at (-2.9cm,-3.6cm) {(b)};
\draw[thick,black!50!] (1.788,-4.15cm) -- (1.789,-5.41cm);
\draw[-latex,thick,black!50!] (1.788,-5.49cm) -- (1.789,-6.45cm);
\end{tikzpicture}
\caption{\label{Fig:freq_sweep} (a) Population transferred to Q1 as function of the frequency shifts applied to both drives when $\theta=1.1~\rm{rad}$ and Q2 is initially in state $\ket{f}$. The dashed lines show a Gaussian fit to determine the frequency shifts $\delta\omega_1$ and $\delta\omega_2$ for optimal population transfer indicated by the dot. (b) Measured values (points) and calibration curves (lines) for the compensation of the $\theta$-dependent cross ac-Stark shifts during two-tone driving. The optimal value for $\theta=1.1~\rm{rad}$ in (a) is indicated by an arrow. The ac-Stark shifts $\Delta_{f0}^{(i)}$ are the fitted curves from Fig.\ \ref{Fig:StarkShift}. In both (a) and (b) frequency shifts are relative to the zero-amplitude offsets $3.196~\rm{GHz}$ and $2.775~\rm{GHz}$ for Q1 and Q2, respectively.}
\end{figure}
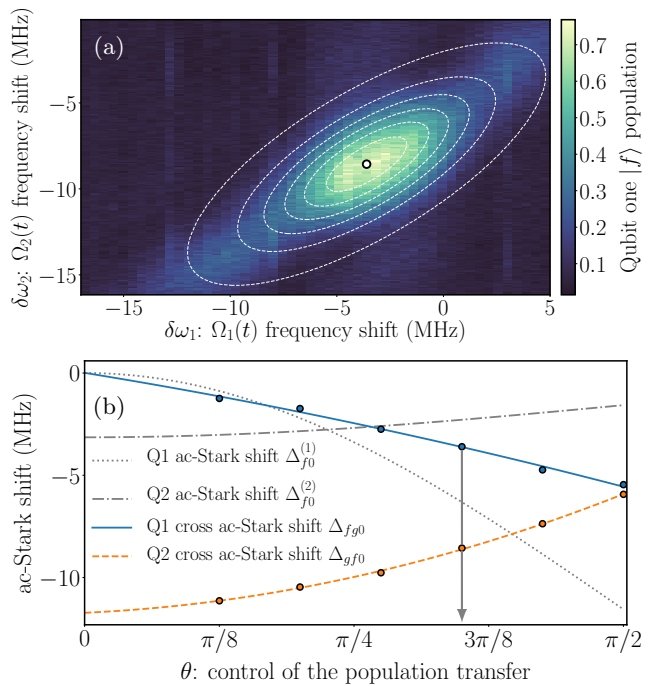
\end{center}

In the next step (ii), the amplitude of each drive (applied separately) is adjusted to produce rotations between the qubit $\ket{f}$ and the resonator $\ket{1}$ state at equal Rabi rates. With a target duration of $T=213~\rm{ns}$ we obtain $\tilde g_1/(2\pi)=4.69(1)~\rm{MHz}$ and $\tilde g_2/(2\pi)=4.72(1)~\rm{MHz}$ for Q1 and Q2, respectively, see Fig.\ \ref{Fig:rates}. For this, the calibration of the ac-Stark shifts of step (i) is used to adjust the drive frequency. 

In step (iii) both drives are simultaneously applied for a duration $T$. To quantify the cross ac-Stark shifts the frequencies of both drives are varied by offsets $\delta\omega_i$ and the population transfer between the qubits is measured. For this, Q2 is prepared in state $\ket{f}$, then both drives are applied, finally the population in Q1 is measured. The population is fitted to a 2D Gaussian function to obtain the frequency offsets $(\Delta_{fg0}, \Delta_{gf0})$ that are equal to the $\delta \omega_i$ values of maximum population transfer and therefore compensate the cross ac-Stark shifts, see Fig.\ \ref{Fig:freq_sweep}(a).

The measured cross ac-Stark shifts differ from the ac-Stark shifts $\Delta_\text{f0}^{(i)}$ obtained when a single tone is applied, see Fig.\ \ref{Fig:freq_sweep}(b). This effect is not observed in the simulations discussed in Sec.\ \ref{Sec:sim}. We therefore attribute the extra shifts to drive-induced crosstalk and/or unwanted qubit-qubit interactions resulting from static capacitive coupling. In the last step (iv) we determine the dependence of the cross ac-Stark shifts on the $\theta$ angles, which is adjusted by the drive amplitude ratio $\lambda_1/\lambda_2$. For this, step (iii) is repeated for different $\theta$. The cross ac-Stark shifts $(\Delta_{fg0}(\theta), \Delta_{gf0}(\theta))$ are  fitted to a second-order function in $\theta$ yielding a calibration curve for each drive, see Fig.\ \ref{Fig:freq_sweep}(b).

The cross ac-Stark shifts acting on the qubits during the simultaneous driving leads  to an additional phase shift of the individual qubit states.  This induces a systematic $\theta$-dependent phase shift on the qubit states that is removed a posteriori in the data so as not to affect fidelity measurements. Alternatively, such phase shifts can also be removed in software \cite{McKay2017}.

\section{Experimental results \label{Sec:experiments}}

\begin{center}
\begin{figure}[!ht]
\includegraphics[width=0.48\textwidth]{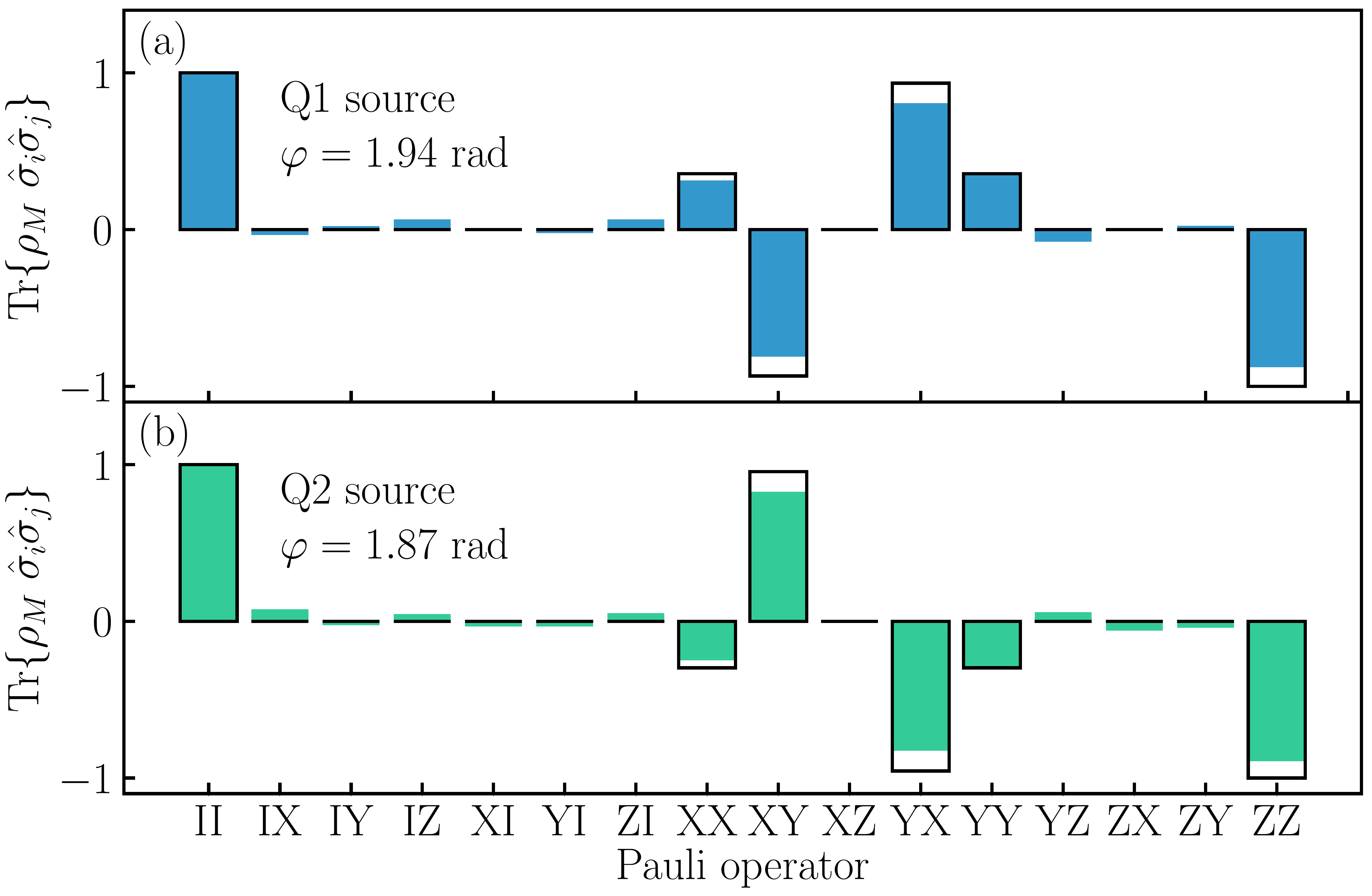}
\caption{\label{Fig:entangled_state} Pauli-operator representation of the entangled state $\ket{\Psi}=(\ket{eg}+e^{i\varphi}\ket{ge})/\sqrt{2}$ with $95.3$\% and $95.2\%$ fidelity when the source qubit is Q1 (a) and Q2 (b), respectively. The solid black lines show the ideal values.}
\end{figure}
\end{center}

To form entangled states as arbitrary superpositions of $\ket{eg}$ and $\ket{ge}$, one qubit (source) is prepared in state $\ket{f}$, the two-tone holonomic operation coherently transfers some population to the $\ket{f}$ state of the other qubit (target), and a final $X_{\pi}^{f\to e}$ pulse maps the qubit $\ket{f}$ populations back to $\ket{e}$. With the chosen rate of $\tilde g/(2\pi)=4.70~\rm{MHz}$ for the holonomic operation and $\theta = \pi/4$, a fully-entangled Bell state $\ket{\Psi}=(\ket{eg} + e^{i\varphi}\ket{ge})/\sqrt{2}$, is formed by a pulse of duration $213~\rm{ns}$. A state fidelity $\mathcal{F}=\text{Tr}\{\sqrt{\sqrt{\rho_\text{T}}\rho_\text{M}\sqrt{\rho_\text{T}}}\} = 95.2\%$ ($95.3\%$) is obtained for Q1 (Q2) as the source qubit with the ideal target state density matrix $\rho_\text{T}=\ket{\Psi}\bra{\Psi}$ and the measured density matrix $\rho_\text{M}$. $\rho_\text{M}$ is determined using state tomography with 1000 repeated single-shot measurements, see Fig.\ \ref{Fig:entangled_state}.

To demonstrate control over the angles $\theta$ and $\phi$, quantum state tomography has been done on $900$ states generated with $30$ different linearly spaced values of $\theta\in[0,\pi/2]$ and $\phi\in[0,2\pi]$. For each state the fidelity $\mathcal{F}$ has been measured with Q1 used as the source qubit. Scaling the drives by $\lambda_1$ and $\lambda_2$ according to Eq.\ (\ref{Eqn:phiTheta}) produces the different $\theta$-angles transferring $\sin(\theta)^2$ of population from the source qubit to the target qubit. Changing the phase $\phi_2$ of the drive on Q2 changes the measured angle $\varphi$.

\begin{center}
\begin{figure}[!ht]
\includegraphics[width=0.48\textwidth]{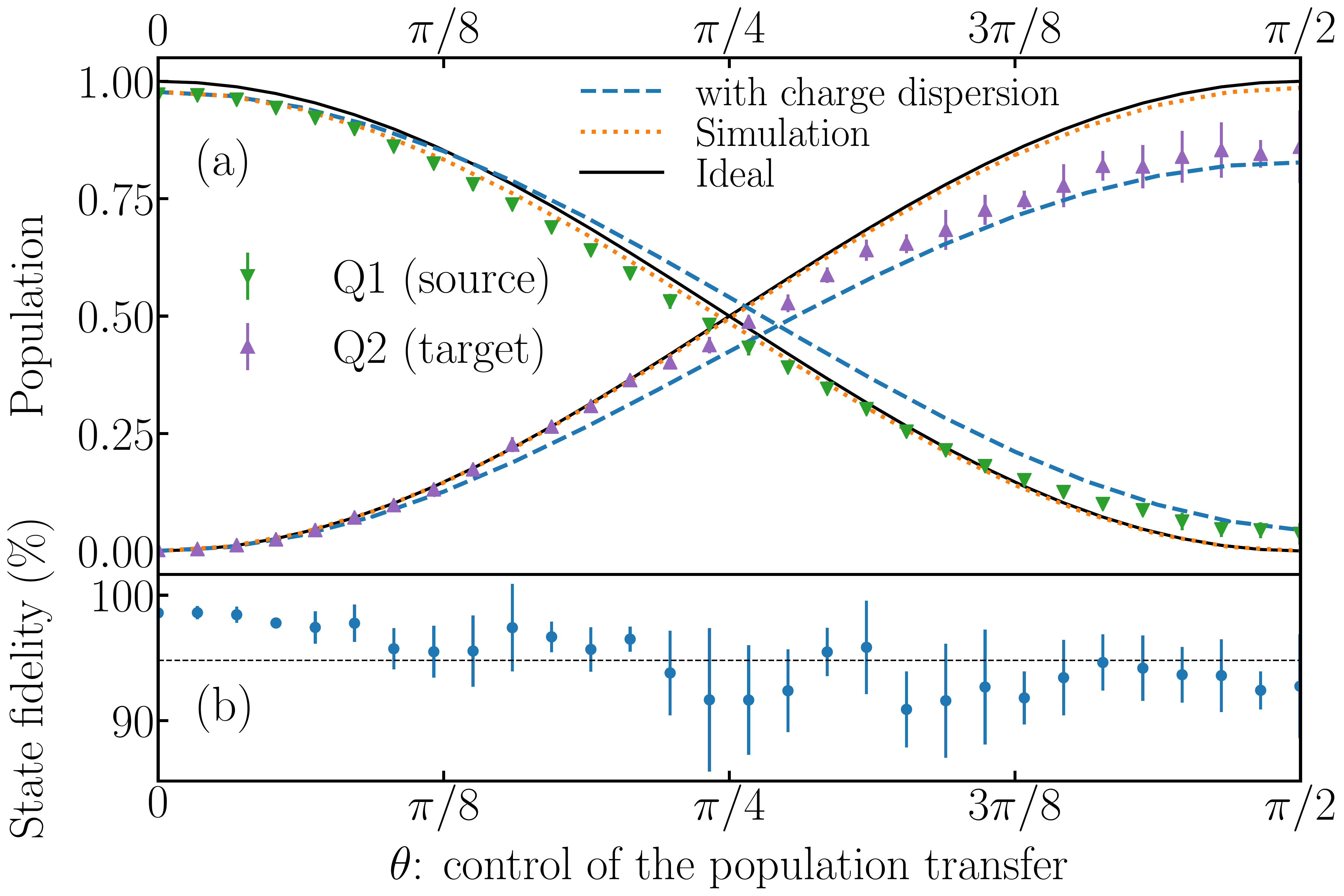}
\caption{\label{Fig:state_transfer} State transfer from Q1 to Q2 as a function of the angle $\theta$. (a) Population in the source (Q1) and target qubit (Q2) extracted from quantum state tomography in the qubit $\{\ket{e}, \ket{g}\}$ subspace. The solid lines indicate ideal values. The dotted lines are a simulation showing the effect of $T_1$ on the ideal population transfer. The dashed lines show a simulation of the effect of charge dispersion and $T_1$. (b) State fidelity after transferring an amount of population controlled by $\theta$. Each point is the state fidelity average over 30 different values of $\phi$. The average state fidelity is 94.8\%.}
\end{figure}
\end{center}

For each value of $\theta$, the populations in the source qubit and target qubit, extracted from the density matrix, are averaged over the different phase values $\phi$. As $\theta$ increases, more population is transferred to the target qubit in good agreement with theory, see Fig.\ \ref{Fig:state_transfer}(a). The state fidelities averaged over $\phi$ are lower for larger $\theta$, see Fig.\ \ref{Fig:state_transfer}(b). We attribute this to the large charge dispersion (determined by Ramsey-type measurements) of $1.8~\rm{MHz}$ and $3.0~\rm{MHz}$ on the $\ket{f}$ states of Q1 and Q2 respectively, as confirmed by numerical simulations discussed in Sec.\ \ref{Sec:sim}.

\begin{center}
\begin{figure}[!ht]
\includegraphics[width=0.47\textwidth]{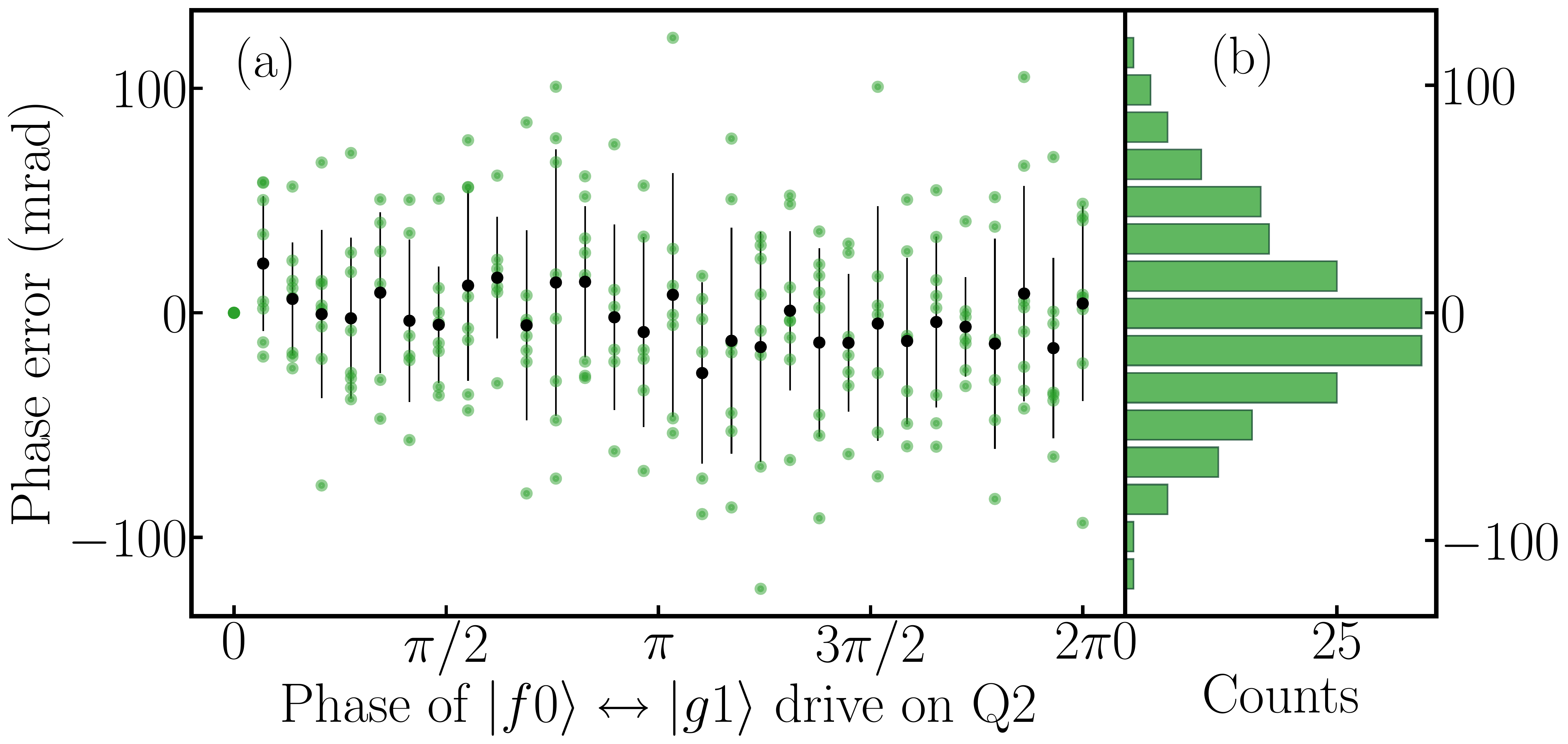}
\caption{\label{Fig:phase_control} (a) Difference between expected and measured phases (between $\ket{ge}$ and $\ket{eg}$) obtained from quantum state tomography as a function of the phase of the drive applied to Q2. Each green point corresponds to different value of $\theta$ and $\phi$. The black dots show the average over $\theta$. The data points at the set phase $\phi=0$ are used to remove the systematic single-qubit phases induced by $\theta$-dependent cross ac-Stark shifts leading to zero dispersion of the first point.
(b) Histogram of the errors, the mean and standard deviation of all points are $-2~\rm{mrad}$ and $42$ mrad, respectively.}
\end{figure}
\end{center}

Control over $\phi$ is demonstrated using the density matrices with $\theta \in [0.19\pi, 0.31\pi]$ from which the phase between $\ket{eg}$ and $\ket{ge}$ can be measured and compared to the expected phase, see Fig.\ \ref{Fig:phase_control}. This $\theta$-range is used since the measured phase between $\ket{eg}$ and $\ket{ge}$ is more accurate when both states have large populations. In 8.3\% of the measurements, a large systematic phase error of $834\pm81$ mrad was observed, which we attribute to charge noise. These data points are not shown in the figure. The difference between the expected and measured relative phase $\varphi$ between the $\ket{eg}$ and $\ket{ge}$ states for $\varphi\neq0$ is $-2\pm42~\rm{mrad}$, see Fig.\ \ref{Fig:phase_control}(b). The small mean shows that $\phi$ can be set with high accuracy. The large standard deviation reflects measurement imperfections of the density matrix and includes the measured timing jitter $\sigma_\text{jitter}=2.6~\rm{ps}$ between the two drives. This corresponds to a $7~\rm{mrad}$ imprecision in $\varphi$ computed from $2\pi\sigma_\text{jitter}(3.196~\text{GHz} - 2.775~\text{GHz})$.

\section{Simulation results \label{Sec:sim}}

The holonomic operation is simulated using QuTiP \cite{Johansson2013a} to understand leading error contributions. We use the Hamiltonian in Eq.\ (\ref{Eqn:H}) with the qubits modeled as anharmonic four-level systems and the resonator modeled as a harmonic three-level system. Experimentally determined frequencies, anharmonicities, coupling strengths and $T_1$ times are taken into account when computing the time evolution using a master equation in Lindblad form. As in the experiment, we assume that one of the qubits starts in the $\ket{f}$ state. The control
pulse is a flat-top Gaussian with $206~\rm{ns}$ long top and $\sigma=3.5~\rm{ns}$ to reflect the effect of the finite bandwidth
of the experimental setup on the square pulse.

\begin{center}
\begin{figure}[!ht]
\includegraphics[width=0.47\textwidth]{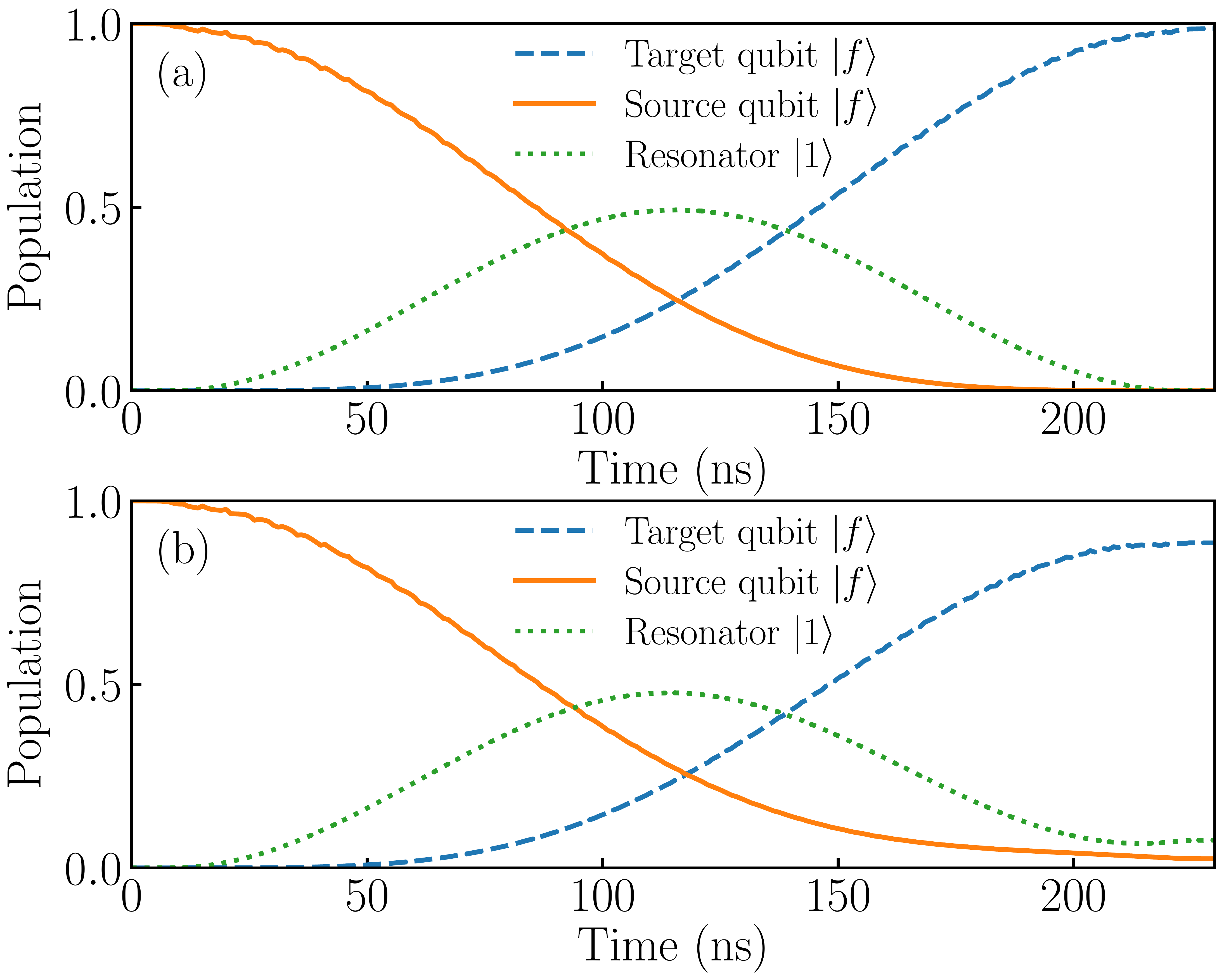}
\caption{\label{Fig:simulation} Simulation of the time evolution when $\theta=\pi/2$ and Q1 is the source qubit. (a) Time evolution assuming the drives are calibrated using the procedure in Sec.\ \ref{Sec:cal}. (b) Simulated time evolution when the drive on Q1 is $0.9~\rm{MHz}$ detuned.}
\end{figure}
\end{center}

The simulated time evolution of a population transfer from Q1 to Q2 shows that the resonator is populated during the holonomic operation and returns to its initial ground state at the end of the pulse, see Fig.\ \ref{Fig:simulation}(a) for an example with $\theta=\pi/2$. Here the maximum population reached in the resonator is 50\%. Table I summarizes simulation results of end populations for different $T_1$ times of the qubits and the resonator. For unitary dynamics ($T_1\to\infty$), the target state is reached with 99.98\% fidelity. If relaxation processes for both qubits and resonator are included, this value decreases to 98.48\%, with some residual population in the $\ket{e}$ states due to the finite $T_1$ time of the  qubits. The short $T_1$ time of the resonator only marginally decreases the amount of population transferred, as can be seen from the fidelity of 99.11\% reached when assuming no relaxation in the resonator but finite $T_1$ times of the qubits. However, the effect of the $1.8~\rm{MHz}$ and $3.0~\rm{MHz}$ charge dispersion on Q1 and Q2, respectively, is detrimental. We simulate this effect by imposing a frequency shift $-0.9~\rm{MHz}\leq\delta\omega_1/(2\pi)\leq0.9~\rm{MHz}$ on the drive of Q1 and $-1.5~\rm{MHz}\leq\delta\omega_2/(2\pi)\leq1.5~\rm{MHz}$ on the drive of Q2. For each pair of frequency shifts on this two-dimensional grid we compute the time dynamics for different $\theta$-angles. We average the final population in each qubit over the two-dimensional grid assuming that the offset charge $n_\text{g}$ due to the electrostatic environment has a uniform probability distribution. Since the qubit frequency is a sinusoid function in $n_\text{g}$ \cite{Schreier2008} the probability distribution function for $\delta\omega_i$ is $f_i(\delta\omega_i)=1/(\pi{\scriptstyle\sqrt{\delta\omega_{i,\text{max}}^2-\delta\omega_i^2}})$ where $\delta\omega_{i,\text{max}}/(2\pi)$ is $0.9~\rm{MHz}$ and $1.5~\rm{MHz}$ for Q1 and Q2, respectively. The population transfer with the simulated charge noise matches well the experimental data, see Fig.\ \ref{Fig:state_transfer}(a). As seen from the last entry in Tab.\ \ref{Tab:pop}, with such large charge dispersion the population is not properly transferred between the qubits and partially remains in the resonator, see Fig.\ \ref{Fig:simulation}(b) for an example in which $(\delta\omega_1/(2\pi), \delta\omega_2/(2\pi))=(0.9,0)~\rm{MHz}$ and $\theta=\pi/2$. We expect that by slightly modifying the charging energies of the transmons, charge noise can be significantly suppressed: for example requiring less than $10~\rm{kHz}$ charge dispersion in the $\ket{f}$ state can be achieved with an $E_C\leq217~\rm{MHz}$ for a $5.0~\rm{GHz}$ qubit \cite{Koch2007}.

\begin{table}
\begin{center}
\caption{\label{Tab:pop} Final populations in the different states of the system rounded to $10^{-4}$ after a flat-top Gaussian pulse with $206~\rm{ns}$ long top and $\sigma=3.5~\rm{ns}$ that swaps the population from $\ket{fg0}$ into $\ket{gf0}$. The numbers are expressed as a percentage. }
\begin{tabular}{x{1.3cm} x{1.3cm} x{1.3cm} x{1.3cm} x{1.3cm} x{1.3cm}}\hline\hline
 \multicolumn{5}{l}{Unitary time evolution} \\ 
 $\ket{gf0}$ &$\ket{gg0}$ &$\ket{eg0}$ &$\ket{ge0}$ &$\ket{gg1}$ & $\ket{fg0}$ \\
 99.98 & 0 & 0 & 0 & 0.02 & 0 \\ \hline
 \multicolumn{5}{l}{Finite $T_1$} \\ 
 $\ket{gf0}$ &$\ket{gg0}$ &$\ket{eg0}$ &$\ket{ge0}$ &$\ket{gg1}$ & $\ket{fg0}$ \\
 98.62 & 0.87 & 0.19 & 0.29 & 0.02 & 0\\ \hline
 \multicolumn{5}{l}{Finite qubit $T_1$, Resonator $T_1\to\infty$} \\
 $\ket{gf0}$ &$\ket{gg0}$ &$\ket{eg0}$ &$\ket{ge0}$ &$\ket{gg1}$ & $\ket{fg0}$ \\
 99.11  & 0.01 & 0.37 & 0.5 & 0.02 & 0\\ \hline
 \multicolumn{5}{l}{Finite $T_1$ with charge dispersion } \\
 $\ket{gf0}$ &$\ket{gg0}$ &$\ket{eg0}$ &$\ket{ge0}$ &$\ket{gg1}$ & $\ket{fg0}$ \\
 82.73 & 0.98 & 0.17 & 0.3 & 11.38 & 4.44 \\ \hline\hline
\end{tabular}
\end{center}
\end{table}

\section{Conclusion and outlook}

We have demonstrated entanglement creation and manipulation of two-qubit states using non-adiabatic holonomic operations. Using square pulses lasting $213~\rm{ns}$ we have created two-qubit states with fidelities above 95\%. This fidelity is also affected by imperfections of the single qubit rotations $X_\alpha^{i\to j}$ with gate fidelities of 98\% as determined by quantum process tomography. Simulations have identified charge dispersion in the qubit $\ket{f}$ states as the main fidelity limitation. With reduced charge dispersion, population transfers with higher fidelity are possible. Further gains in fidelity can be obtained by increasing the $T_1$ time of the resonator, e.g. by optimizing its geometry \cite{Wenner2011}. In the future, the holonomic operation presented here may be extended to a two-qubit non-Abelian non-adiabatic gate \cite{Hong2017}. For this to be possible, the 1-2 excitation manifold $\{\ket{gg1},\ket{fg0},\ket{gf0}\}$ must be well separated from the transitions in the 3-4 excitation manifold $\{\ket{fg1},\ket{gf1},\ket{ff0}\}$ to avoid residual driving of these transitions. In our sample, the separation between these manifolds was $5~\rm{MHz}$. A larger separation will permit the operation of Eq.\ (\ref{Eqn:U}) as a SWAP-type gate that acts on the full computational subspace without population loss. A larger separation can be achieved by increasing the dispersive shift between the coupling resonator and the qubits. Alternatively, it may be possible to use more complex pulse shapes obtained by optimal control methods \cite{Glaser2015, Machnes2015} applied to superconducting qubits \cite{Egger2013a, Egger2014}.

\section{Acknowledgments}
We would like to thank S. Gasparinetti, A. Wallraff and S. Machnes for useful discussions and R. Heller and H. Steinauer for electronics support. This work was supported by the IARPA LogiQ program under contract W911NF-16-1-0114-FE and the ARO under contract W911NF-14-1-0124. D.E. and S.F. acknowledge support by the Swiss National Science Foundation (SNF, Project 150046).

\appendix

\section{Dynamics \label{Sec:parallel}}

Here we show that the evolution in the two-qubit subspace governed by the Hamiltonian $\hat H'_\text{eff}$ from Eq.\ (\ref{Heff_no_stark}) is purely geometric by following a parallel transport. Consider two orthonormal vectors that initially span the qubit subspace $\{\ket{fg0},\ket{gf0}\}$
\begin{align} \notag
\ket{\psi_1(0)}&=\alpha\ket{fg0}+\beta\ket{gf0} \\ \notag
\ket{\psi_2(0)}&=\beta^*\ket{fg0}-\alpha^*\ket{gf0},
\end{align}
where $\alpha,\beta\in \mathbb{C}$ and $|\alpha|^2+|\beta|^2=1$. We define the initial subspace $S_0={\rm span}\{\ket{\psi_1(0)}, \ket{\psi_2(0)}\}$. Under the action of $\hat H'_\text{eff}$ this subspace evolves into $S_t={\rm span}\{\ket{\psi_1(t)}, \ket{\psi_2(t)}\}$. The time evolution satisfies the parallel transport condition 
\begin{align} \label{Eq:parallel}
\braket{\psi_j(t)|\hat H'_\text{eff}(t)|\psi_k(t)}=0
\end{align} 
which can be transformed into $\braket{\psi_j(t)|\dot\psi_k(t)}=0$ using the Schrodinger equation. This implies that an infinitesimally small time-step ${\rm d}t$ evolves the vector $\ket{\psi_k(t)}$ along a direction which is perpendicular to all vectors in $S_t$, i.~e. there are no transitions between the vectors in $S_t$ during the time evolution. This means that the resulting non-Abelian geometric phase has no dynamical contributions \cite{Sjoqvist2015}.

Now we show that Eq.\ (\ref{Eq:parallel}) holds when the Hamiltonian $\hat H'_\text{eff}$ is given by Eq.\ (\ref{Heff_no_stark}). We write $\hat H'_\text{eff}(t)=\tilde g(t)\hat A$, where
\begin{align} \notag
\hat A = \begin{pmatrix}
0 & \lambda_1 & 0 \\ \lambda_1^* & 0 & \lambda_2 \\
0 & \lambda_2^* & 0
\end{pmatrix}
\end{align}
in the basis $\ket{fg0}=(1, 0, 0)^T$, $\ket{gg1}=(0, 1, 0)^T$ and $\ket{gf0}=(0, 0, 1)^T$. The time evolution operator is
\begin{align} \notag
\hat U(t)=\mathcal{T}\exp\left(-i\int_0^t \tilde g(t')\hat A{\rm d}t\right)=\exp\left(-iG(t)\hat A\right)
\end{align}
where $G=\int_0^t \tilde{g}(t') {\rm d}t'$ and $\mathcal T$ denotes time-ordering. The last  equation follows from the time-independence of $\hat A$ resulting in $\hat H'_\text{eff}$ commuting with itself at all times. The left-hand side of Eq.\ (\ref{Eq:parallel}) can be written as $\tilde g(t)\braket{\psi_j(0)|\hat U(t)^\dagger\hat A\hat U(t)|\psi_k(0)}$. Since $[\hat U,\hat A]=0$ the term $\hat U^\dagger\hat A\hat U$ reduces to $\hat A$. The action of $\hat A$ on $\ket{\psi_1(0)}$ and $\ket{\psi_2(0)}$ yields a vector proportional to $\ket{gg1}$ which is perpendicular to $\ket{\psi_j(0)}$ for $j=1,2$. The parallel transport condition is thus satisfied $\forall~t$.

\section{Cyclical evolution \label{Sec:dynamics}}
Since $|\lambda_1|^2+|\lambda_2|^2=1$, $\hat A$ satisfies $\hat A^{3}=\hat A$ resulting in $\hat A^{2n+1}=\hat A$ and $\hat A^{2n}=\hat A^2$ for $n\geq1$ where 
\begin{align} \notag
\hat A^2 = \begin{pmatrix}
|\lambda_1|^2 & 0 & \lambda_1\lambda_2 \\ 
0 & 1 & 0 \\
\lambda_1^*\lambda_2^* & 0 & |\lambda_2|^2
\end{pmatrix}.
\end{align}
Using the Taylor expansion for the exponential, cosine and sine functions the time evolution operator is
\begin{align} \notag
\hat U(t)=&~I+\hat A^2\{\cos[G(t)]-1\}-i\hat A\sin[G(t)].
\end{align}
When $G(T)=\pi$ the time evolution operator reads $\hat U(\tau)=I-2\hat A^2$. The evolution is thus cyclical, i.~e. $S_T=S_0$, since $\hat A^2$ does not mix the subspace $\{\ket{fg0},\ket{gf0}\}$ and $\{\ket{gg1}\}$. By introducing $-e^{i\phi}\tan(\theta/2)=\lambda_1/\lambda_2$ one recovers the time evolution operator of Eq.\ (\ref{Eqn:U}) in the main text.

\bibliography{../IBMRefDB}

\end{document}